\documentclass[preprints,article,accept,moreauthors,pdftex]{Definitions/mdpi}

\firstpage{1} 
\pubvolume{xx}
\issuenum{x}
\articlenumber{x}
\pubyear{x}
\copyrightyear{x}
\history{~}

\usepackage{todonotes}

\Title{Random Walk on a Rough Surface: Renormalization Group Analysis of a Simple Model}

\Author{Nikolay~V.~Antonov,$^{1,2}$
Nikolay~M.~Gulitskiy,$^{1}$
Polina~I.~Kakin$^{1}$ 
and
Dmitriy~A.~Kerbitskiy$^{1}$}

\address{
	$^{1}$ \quad Department of Physics,  
	Saint Petersburg State University, Universitetskaya nab.~7/9, St. Petersburg, 199034, Russia;\\
				$^2$  \quad 
	 N.N. Bogoliubov Laboratory of Theoretical Physics, Joint Institute for Nuclear Research, Dubna, 141980, Moscow Region, Russia.	}

\abstract{The field theoretic renormalization group is applied to a simple model of random walk on a rough fluctuating surface. We consider the Fokker--Planck equation for a particle in a uniform gravitational field. The surface is modelled by the generalized Edwards--Wilkinson linear stochastic equation for the height field. The full stochastic model is reformulated as a multiplicatively renormalizable field theory, which allows for application of the standard renormalization theory. The renormalization group equations have several fixed points that correspond to possible scaling regimes in the infrared range (long times, large distances); all the critical dimensions are found exactly. As an example, the spreading law for particle's cloud is derived. It has the form $R^2(t)\simeq t^{2/\Delta_{\omega}}$ with the exactly known critical dimension of frequency $\Delta_{\omega}$ and, in general, differs from the standard expression $R^2(t)\simeq t$ for ordinary random walk.}


\begin{document}

\section{Introduction}
\label{Intro}

Over decades, stochastic growth processes, kinetic roughening phenomena and fluctuating surfaces or interfaces have been attracting constant attention. The most prominent examples include deposition of a substance on a surface and the growth of the corresponding phase boundary; propagation of flame, smoke, and solidification fronts; growth of vicinal surfaces and bacterial colonies; erosion of landscapes and seabed profiles; molecular beam epitaxy and many others; see \cite{EW}--\cite{Xia} and references therein.

Another vast area of research is that of diffusion and random walks in random environment such as disordered, inhomogeneous, porous or turbulent media; see, e.g. \cite{Walks}--\cite{Walks3}.

In this paper, we study a simple model of a random walk on a rough fluctuating surface. We consider the Fokker--Planck equation for a particle in a uniform gravitational field. The surface is modelled by the generalized Edwards--Wilkinson linear stochastic equation for the height field \cite{EW}. The generalized model involves two arbitrary exponents: $\varepsilon$ and $\eta$, related to the spectrum and the dispersion law of the height field, respectively. Detailed description of the model and its relation to various special cases is given in Sec.~\ref{Model}.

Using the general Martin--Siggia--Rose--De~Dominicis--Janssen theorem, the original stochastic problem is reformulated as a certain field theoretic model. This allows one to apply the well-developed formalism of Feynman diagrammatic techniques, renormalization theory and renormalization group (RG). The model is shown to be multiplicatively renormalizable, so that the RG equation can be derived in an standard way. The corresponding renormalization constants and the RG functions (anomalous dimensions and $\beta$ functions) are explicitly calculated in the leading one-loop order of the RG perturbation theory. These issues are discussed in Secs.~\ref{QFT} and~\ref{RGE}.

The RG equations have two Gaussian (free) fixed points and two nontrivial ones. 
Those points are infrared (IR) attractive depending on the values of the parameters $\varepsilon$ and $\eta$, which implies the existence of scaling (self-similar) asymptotic regimes in the IR range (long times and large distances) for various response and correlation functions of the model (Sec.~\ref{RGE}). The critical dimensions for those regimes are found exactly as functions of $\varepsilon$ and $\eta$. As an indicative
application, the time dependence of the mean-square radius of a cloud of randomly walking particles is obtained (Sec.~\ref{Scaling}). It is described by a power law with the exponent that depends on the fixed point, is known exactly as a function of $\varepsilon$ and $\eta$ and, for nontrivial points, differs from the ordinary random walk: $R^2(t)\simeq t$.

Some implications and possible generalizations are discussed in Sec.~\ref{Conc}.  

\section{Description of the model}
\label{Model}

We consider the following stochastic problem for a random walk:
\begin{equation}
\partial_t x_i = F_i({\bf x}) + \zeta_i, \quad 
\langle \zeta_i(t)\zeta_j(t') \rangle_{\zeta} =2\nu_0\delta(t-t').
\label{RW}
\end{equation}
Here ${\bf x}(t) =\{x_i(t)\}$ is the coordinate of the particle, $i=1\dots d$, where $d$ is an arbitrary (for generality) dimension of the ${\bf x}$ space, $\zeta_i=\zeta_i(t)$ is a Gaussian random noise with zero mean and a given pair correlation function, $\nu_0>0$
is the diffusion coefficient, and $F$ is an external ``drift'' force.\footnote{Here and below, the subscript 0 refers to bare parameters which will be renormalized in the following.}
The probability distribution function $P(t,{\bf x})$ satisfies the (deterministic) 
Fokker--Planck equation
\begin{equation}
\left\{ \partial_t + \partial_i (F_i - \nu_0\partial_i)\right\}\, P(t,{\bf x})=0 
\label{FPE}
\end{equation}
(here and below, summation over repeated indices is implied).
For a particle in a constant gravitational field one has 
\begin{equation}
F_i= - \lambda_0 \partial_i h,
\label{gravi}
\end{equation}
where $\lambda_0 =mg$, $g$ is the gravitational acceleration,
$m$ is the particle's  mass, and $h$ is the height of its location.

The simplest model of a surface roughening, proposed within the context of landscape erosion, is the one due to Edwards and Wilkinson \cite{EW}. In the continuous formulation, it is described by the diffusion-type stochastic equation for the height field $h=h(t,{\bf x})$: 
\begin{equation}
\{ \partial_t - \kappa_0 \partial^2 \}\, h(t,{\bf x})= f(t,{\bf x}),
\label{EW}
\end{equation}
where $\kappa_0>0$ is (a kind of) surface tension coefficient, $\partial^2=\partial_i\partial_i$ is the Laplace operator and $f$ is a Gaussian random noise with zero mean and a given pair correlation function.
The most popular choices are the white noise
\begin{equation}
\langle f(t,{\bf x})f(t',{\bf x'})\rangle_{f} = D_0 \delta(t-t')
\delta({\bf x}-{\bf x'})
\label{White}
\end{equation}
with the positive amplitude $D_0>0$, and the quenched noise; the simplified version of the latter is 
\begin{equation}
\langle f(t,{\bf x})f(t',{\bf x'})\rangle_{f} = D_0 \delta({\bf x}-{\bf x'}).
\label{quench}
\end{equation}
In this paper, we consider a generalized equation 
\begin{equation}
\{ \partial_t + \kappa_0 k^{2-\eta}\}\, h(t,{\bf x})= f(t,{\bf x}),
\label{EWG}
\end{equation}
written here in the symbolic notation with $k$ being the wave number,
\footnote{Detailed discussion of fractional derivatives can be found in \cite{WalksX}.} while the correlation function is taken in a power-like form:
\begin{equation}
\langle f(t,{\bf x})f(t',{\bf x'})\rangle_f = D_0  \delta(t-t')\,
\int \frac{d{\bf k}}{(2\pi)^d} \, k^{2-d-y}\,  \exp \{ i{\bf k} ({\bf x}-{\bf x'}) \}.
\label{White2}
\end{equation}
Here $\eta$ and $y$ are arbitrary exponents and $d$ is the dimension of space. Clearly, the choice $\eta=0$, $2-d-y=0$ corresponds to the model (\ref{EW}), (\ref{White}); as we will see, the model (\ref{EW}), (\ref{quench}) can also be obtained from (\ref{EWG}), (\ref{White2}). 

The choice $\eta\ne0$ can be justified by the ideas of self-organized criticality (SOC) according to which the evolution of a sandpile surface is not an ordinary diffusion-type process but involves several discrete steps: expectation period, reaching a threshold and avalanche; see, e.g.~\cite{Bak3}.  

For a linear stochastic equation with a Gaussian additive random noise, the field $h$ is also a Gaussian field defined by its pair correlation function. For the model (\ref{EWG}), (\ref{White2}), the latter has the following form in the Fourier ($\omega$--${\bf k}$) representation
\begin{equation}
D_h(\omega,{k}) = \frac{D_0 \, k^{2-d-y}} {\omega^2 + [\kappa_0 k^{2-\eta}]^2}
= \frac{g_0 u_0 \nu_0^3 \, k^{2-d-\eta-\varepsilon}} {\omega^2 + [u_0\nu_0 k^{2-\eta}]^2}.
\label{Dh}
\end{equation}
In the second relation we introduced the new variables: the exponent $\varepsilon$ and the amplitudes $g_0$, $u_0$, defined by the relations 
\begin{equation}
\varepsilon = y-\eta, \quad \kappa_0 =u_0\nu_0, \quad D_0 = g_0 u_0 \nu_0^3.
\label{Defin}
\end{equation}
They are convenient, in particular, because the equal-time correlation function
\begin{equation}
D_h(k)= \int \frac{d\omega}{2\pi} D(\omega,k) \propto g_0\, \nu_0^2\, k^{-d-\varepsilon}
\label{Equal}
\end{equation}
involves the parameters $g_0$, $\varepsilon$, while the dispersion law
$\omega(k) \propto u_0 \nu_0 k^{2-\eta}$ is expressed only via $u_0$, $\eta$.

The model (\ref{Dh}) includes two special cases interesting on their own. In the limit $u_0 \to \infty$ and $g_0' = g_0/u_0$ fixed, the function $D(\omega, k)$ becomes independent of the frequency $\omega$, and the field $h(t,{\bf x})$ becomes white in time. Indeed, one obtains in the ($t$--${\bf k}$) representation 
\begin{equation}
D(t-t', k) = \delta(t-t')\, g_0'\, \nu_0^2\, k^{-2-d-\varepsilon+\eta}.
\label{Rapid}
\end{equation}
In the limit $u_0 \to 0$ and $g_0$ fixed, the function $D_h(k)$ in (\ref{Equal}) remains finite, so that (\ref{Dh}) tends to  
\begin{equation}
D_h(\omega,{k}) = \pi\delta(\omega)\,   g_0\, \nu_0^2\,  k^{-d-\varepsilon},
\label{Frozen}
\end{equation}
which corresponds to the time-independent (quenched or frozen) field $h$. Surprisingly enough, for $\varepsilon = 4-d$, this reproduces the model (\ref{EW}), (\ref{quench}) where one has $D_h \propto \delta(\omega)/ k^4$.

Substituting the gravitational force (\ref{gravi}) with the random height field from 
(\ref{EWG}), (\ref{White2}) into the Fokker-Planck equation (\ref{FPE}) turns the latter into a stochastic equation in its own right.

This completes formulation of the problem.

\section{Field theoretic formulation and renormalization of the model}
\label{QFT}

According to the general theorem (see, e.g. Sec.~5.3 in monograph \cite{Vasiliev}), the full stochastic problem (\ref{FPE}), (\ref{gravi}), (\ref{EWG}), (\ref{White2}) is equivalent to the field theoretic model for the doubled set of fields $\Phi = \{\theta',h', \theta, h\}$ with the De Dominicis--Janssen action functional:
\begin{eqnarray}
{\cal S}(\Phi) = \theta' \left[-\partial_t \theta + \nu_0 \partial^2\theta + \lambda_0 \partial_i (\theta \partial_i h) \right] + {\cal S}_h(h',h), 
\label{Action}
\\
{\cal S}_h(h',h) = \frac{1}{2} h'D_f h'+ h'\left[-\partial_t + \kappa_0 k^{2-\eta} \right] h.
\label{Actionh}
\end{eqnarray}
Here $D_f$ is the correlator (\ref{White2}),
$\theta$ is the density field, $h$ is the height field and $\theta'$, $h'$ are the corresponding Martin--Siggia--Rose response fields; all the needed integrations over their arguments $x=\{t,{\bf x}\}$ and summations over repeated indices are implied. The field theoretic formulation means that various correlation and response functions of the original stochastic problem are represented by functional averages with the weight $\exp {\cal S}(\Phi)$. The field $h'$ can easily be removed by Gaussian integration, then 
${\cal S}_h(h',h)$ would be replaced with ${\cal S}_h(h)= - h D_h^{-1} h/2$
with $D_h$ from (\ref{Dh}), but the expanded representation (\ref{Actionh}) is more convenient for the renormalization purposes.
The constant $\lambda_0$ can be removed by rescaling of the fields $h,h'$ and other parameters. Thus, in the following, with no loss of generality, we set $\lambda_0=1$.

The model (\ref{Action}), (\ref{Actionh}) corresponds to Feynman diagrammatic technique with bare propagators $\langle \theta'\theta \rangle_0$, $\langle hh \rangle_0$, $\langle h'h \rangle_0$ (the latter does not enter into relevant diagrams) and the only vertex $\theta'\partial_i (\theta \partial_i h)$.

It is well known that analysis of ultraviolet (UV) divergences is based on analysis of canonical dimensions, see, e.g. \cite{Vasiliev} (Secs.~1.15, 1.16).
In contrast to conventional static models, dynamic ones have two independent scales: a time scale $[T]$ and a spatial scale $[L]$; see \cite{Vasiliev} (Secs.~1.17, 5.14).
Thus, the canonical dimension of any quantity $F$ (a field or a parameter) is determined by two numbers: the frequency dimension $d_F^\omega$ and the momentum dimension $d_F^k$: 
\begin{equation} 
\left[F\right] \sim \left[T\right]^{-d_F^\omega} \left[L\right]^{-d_F^k}.
\nonumber
\end{equation}
The dimensions are found from obvious normalization conditions
\begin{equation}
	d_{\bf k}^k = -d_{\bf x}^k = 1, \quad d_{\bf k}^\omega = d_{\bf x}^\omega = 0, \quad d_\omega^k = d_t^k = 0, \quad d_{\omega}^{\omega}=-d_t^{\omega}=1
	\nonumber
\end{equation}
and from the requirement that all terms in the action functional be dimensionless with respect to both the canonical dimensions separately.
The total canonical dimension  is defined as 
$d_{F} = d_{F}^{k} + 2d_{F}^{\omega}$ (the coefficient $2$ follows from the relation $\partial_t \propto {\bf \partial}^2$ in the free theory). In the renormalization procedure, $d_{F}$ plays the same role as the conventional (momentum) dimension does in static models; see~Sec.~5.14 in~\cite{Vasiliev}.

Canonical dimensions of all the fields and parameters of our model are given in Table~\ref{t1}. It also involves renormalized parameters (without subscript  ``0'') and the reference mass $\mu$, an additional parameter of the renormalized theory; they all will appear later on. 

Note that for the fields $\theta'$, $\theta$ all these dimensions can be unambiguously defined only for the product $\theta'\theta$. Formally, this follows from the invariance of the action functional (\ref{Action}) under the dilatation 
$\theta'\to\lambda\theta'$, $\theta\to\lambda^{-1}\theta$.

\begin{center}
\begin{table}[h!]
\centering
\caption{Canonical dimensions for the action functional~(\ref{Action}), (\ref{Actionh}).}
\label{canonical dimensions}
\begin{tabular}{|c||c|c|c|c|c|c|c|c|}
 \hline
$F$&  $\theta'\theta$ & $h'$ & $h$ &$\nu_0$, $\nu$ & $g_{0}$ & $u_0$
& $g$, $u$  &{$\mu$,$m$}\\
\hline\hline
$d^{k}_F$ & $d$ & $d+2$ & $-2$ & $-2$ & $\varepsilon$ & $\eta$ & $0$ & $1$\\
 \hline
$d^{\omega}_F$ & $0$ &$-1$ &$1$& $1$ & $0$& $0$ & $0$ & $0$\\
 \hline
$d_F$ & $d$ & $d$ & $0$ & $0$ & $\varepsilon$ & $\eta$ & $0$ & $1$\\ 
 \hline
\end{tabular}
\label{t1}
\end{table}
\end{center}

As can be seen from Table~\ref{t1}, the model becomes logarithmic (both coupling constants $g_0$, $u_0$ become dimensionless) for $\eta=y=0$ (or equivalently for  $\varepsilon=y=0$) and arbitrary $d$.\footnote{Although $u_0$ is not an expansion parameter in perturbation theory, its renormalized counterpart is dimensionless, enters into renormalization constants and RG functions and should be treated on equal footing with $g_0$. We also recall that $\lambda_0=1$.} According to general strategy of renormalization, the exponents $\eta$, $y$ or $\varepsilon$  that ``measure'' deviation from  logarithmicity, should be treated as formal small parameters of the same order.
The UV divergences manifest themselves as singularities at $y\to 0$, {\it etc.} in the correlation functions; in the one-loop approximation, they have the form of simple poles. 

The total canonical dimension of a certain 1-irreducible Green's functions is given by 
\begin{equation}
d_{\Gamma} = (d+2) - \sum_{\Phi} d_{\Phi} N_{\Phi},
 \label{index}
\end{equation}
where $N_{\Phi}$ are the numbers of the fields $\Phi = \{\theta',h', \theta, h\}$  entering the Green's function and $d_{\Phi}$ are their total canonical dimensions.

The formal index of divergence $\delta_{\Gamma}$ is the total dimension of the Green's function in the logarithmic theory ($y=\eta=0$), that is,  $\delta_{\Gamma}=d_{\Gamma}|_{y=\eta=0}$.
Superficial UV divergences, whose removal requires introducing counterterms, can be present in the Green's function $\Gamma$ if $\delta_{\Gamma}$ is a non-negative integer. 

When analyzing the divergences in the model (\ref{Action}), (\ref{Actionh}), the following additional considerations should be taken into account; see, e.g.
\cite{Vasiliev} (Sec.~5.15) and \cite{RedBook} (Sec.~1.4). 

(i) For any dynamic model of this type, all the 1-irreducible functions without the response fields contain closed circuits of retarded propagators
$\langle\theta\theta'\rangle_0$ and vanish. Thus, it is sufficient to consider the functions with $N_{\theta'}+N_{h'} \ge 1$.

(ii) For all non-vanishing functions, $N_{\theta'}=N_{\theta}$ (otherwise no diagrams can be constructed). Formally, this is  a consequence of the invariance of the action functional (\ref{Action}) with respect to dilatation $\theta'\to\lambda\theta'$, $\theta\to\lambda^{-1}\theta$.

(iii) Using integration by parts, one derivative in the vertex can be moved onto the field $\theta'$, i.e. $\theta'\partial_i (\theta \partial_i h) \simeq -(\partial_i \theta')(\partial_i h)\theta$.
Thus, in any 1-irreducible diagram, each external field $\theta'$ or $h'$, 
``releases'' the  external momentum, and the real index of divergence decreases by the corresponding number of units, i.e. $\delta'=\delta-N_{\theta'}-N_{h}$.
Furthermore, these fields enter the counterterms only in the form of spatial gradients.
This observation excludes the counterterms $\theta'\partial_t \theta$
and $(\theta'\theta)^2$, the latter allowed by the formal index for $d\le 2$.

(iv) It is clear that the fields $\theta'$, $\theta$ do not affect the statistics 
of the field $h$. In the field theoretic terms, this ``passivity'' means that any
1-irreducible Green's function with $N_{\theta'}=0$, $N_{\theta}>0$ and $N_{h}+N_{h'}>0$ vanishes: no corresponding diagrams can be constructed.

Taking into account these considerations one obtains:
\begin{equation}
    \delta= (d+2) -d (N_{\theta}+N_{h'}), \quad
    \delta'= (d+2) -(d+1) N_{\theta} - N_{h} - d N_{h'}
\label{deltas}    
\end{equation}
(we recall that $N_{\theta'}=N_{\theta}$, so that only $N_{\theta}$ is indicated).

Then the straightforward analysis shows that the superficial divergences in our model are present only in the 1-irreducible functions $\langle \theta'\theta\rangle$ and $\langle \theta'\theta h\rangle$, and the corresponding counterterms necessarily contract to the forms $\theta'\partial^2 \theta$ ($\delta=2$, $\delta'=1$) and
$(\partial_i \theta')(\partial_i h)\theta$  ($\delta=2$, $\delta'=0$). 
Such terms are already present in the action (\ref{Action}), which means that our model (\ref{Action}), (\ref{Actionh}) is multiplicatively renormalizable with only two independent renormalization constants $Z_{1}$ and $Z_{2}$.

The renormalized action has the form
\begin{eqnarray}
{\cal S}_R(\Phi) = \theta' \left[-\partial_t \theta +  Z_1 \nu\partial^2\theta + Z_2 \partial_i (\theta \partial_i h) \right] + {\cal S}_{hR}(h',h), 
\label{Raction}
\end{eqnarray}
which is naturally reproduced as renormalization of the field $h$ and the coefficient $\nu_0$; no renormalization of the product $\theta\theta'$ is needed:
\begin{eqnarray}
 \nu_0 = \nu Z_{\nu}, \quad Z_{\nu}=Z_1, \quad Z_{h}=Z_2, \quad Z_{\theta\theta'}=1.
\label{R1}
\end{eqnarray}
The functional (\ref{Actionh}) is not renormalized, ${\cal S}_{hR}(h',h)={\cal S}_{h}(h',h)$,
but it should be expressed in renormalized variables taking into account Eqs. (\ref{White2}) and (\ref{Dh}): 
\begin{eqnarray}
g_0 = g\mu^{y}Z_g, \quad u_0 = u\mu^{\eta}Z_u,\quad \kappa_0 = \kappa Z_{\kappa},
\label{Id}
\end{eqnarray}
where the renormalization mass $\mu$ is introduced so that renormalized couplings $g$ and $u$ are completely dimensionless.
Then it follows from the absence of renormalization of ${\cal S}_{h}$ that
\begin{eqnarray}
Z_h Z_{h'}=1, \quad Z_{h'}^2 Z_g Z_u Z_{\nu}^3=1, \quad Z_u Z_{\nu}= Z_{\kappa}=1.
\label{Id1}
\end{eqnarray}
Along with (\ref{R1}) this finally gives the following relations:
\begin{eqnarray}
Z_g = Z_2^2 Z_1^{-1}, \quad Z_u =Z_1^{-1}, \quad Z_{\nu} =Z_1.
\label{Rach}
\end{eqnarray}

We calculated the renormalization constants $Z_{1}$ and $Z_{2}$ in the leading one-loop approximation (the first order of the perturbative expansion in $g$). It is sufficient to find them for $\eta=0$, because the
anomalous dimensions in the minimal subtraction (MS) renormalization scheme are independent of the parameters like $\eta$ and $y$, while the 
exponent $y$ alone provides UV regularization. Then one obtains:
\begin{eqnarray}
Z_1 =1 -\frac{g}{y}\,\frac{ C_d}{2d}\, \frac{(u-1)}{(u+1)^2}\,, \quad
Z_2 =1 + \frac{g}{y}\,\frac{ C_d}{2d}\, \frac{1}{(u+1)^2},
 \label{ZZ}
\end{eqnarray}
with the higher-order corrections in $g$.
Here $C_d = S_d/(2\pi)^d$, $S_d= 2\pi^{d/2}/\Gamma(d/2)$ is the surface area of the unit sphere in $d$-dimensional space. It is convenient to absorb overall factors into the coupling constant $g$, which gives
\begin{eqnarray}
Z_1 =1 -\frac{g }{y}\, \frac{(u-1)}{(u+1)^2}\, , \quad
Z_2 =1 + \frac{g }{y}\, \frac{1}{(u+1)^2}.
 \label{ZZ1}
\end{eqnarray}
For $\eta\ne0$, the expressions (\ref{ZZ}), (\ref{ZZ1}) would be infinite sums; see, e.g. \cite{Ant1,Ant2}.

\section{RG equations, RG functions, and fixed points}
\label{RGE}

Since our model is multiplicatively renormalizable, the corresponding RG equations are derived in a standard fashion. In particular, for a certain renormalized (full or connected) Green's function $W^R$ the RG equation reads:
\begin{equation}
\left\{ {\cal D}_{\mu} + \beta_g \partial_g + \beta_u \partial_u
 - \gamma_{\nu} {\cal D}_{\nu} - \sum_{\Phi} N_{\Phi} \gamma_{\Phi} 
 \right\}\, W^R(g,u,\nu,\mu;\dots)=0.
\label{RGEq}
\end{equation}
Here the ellipsis stands for other variables (times and coordinates or frequencies and momenta),  $\partial_x = \partial/\partial x$,
 ${\cal D}_x = x \partial_x$ for any variable $x$  and the sum runs over all fields $\Phi = \{\theta',h', \theta, h\}$.

The coefficients in the RG differential operator (\ref{RGEq}) --
the anomalous dimensions $\gamma$ and the $\beta$ functions -- are defined as:
\begin{equation}
\gamma_{\alpha}= \widetilde{\cal D}_{\mu}\, \ln Z_{\alpha} \quad
{\rm for\ any\ }\,\alpha, \quad 
\beta_g = \widetilde{\cal D}_{\mu}\,g, \quad
\beta_u = \widetilde{\cal D}_{\mu}\, u,
\label{RGF}
\end{equation}
where $\widetilde{\cal D}_{\mu}$ is the differential operation ${\cal D}_{\mu}$ at fixed bare (unrenormalized) parameters; see, e.g. Secs. 1.~24, 1.~25  in the monograph \cite{Vasiliev}.

 From (\ref{R1})--(\ref{Rach}) and definitions (\ref{RGF}) it follows that 
\begin{eqnarray}
\gamma_{\theta\theta'} = 0, \quad
\gamma_h = - \gamma_{h'} = \gamma_2, \quad 
\gamma_g = 2 \gamma_2 - \gamma_1, \quad 
\gamma_u = - \gamma_{\nu} = - \gamma_1, 
\label{Gammas3}  \\
\beta_g = g [-\varepsilon-\gamma_g], \quad 
\beta_u= u [-\eta-\gamma_u].
\label{Gammas4}    
\end{eqnarray}
From (\ref{Gammas4}) and the one-loop result (\ref{ZZ1}) one obtains:
\begin{equation}
\gamma_1 = g \,\frac{u-1}{(u+1)^2}, \quad \gamma_2 =- g\,\frac{1}{(u+1)^2}\,,
\label{Gammas2}    
\end{equation}
\begin{eqnarray}
\beta_g = g \left[ - \varepsilon + g\,\frac{2  u}{(u+1)^2}\right]\,, \qquad
\beta_u = u \left[ - \eta + g\,\frac{u-1}{(u+1)^2}\right]\,,
\label{Betau}    
\end{eqnarray}
with the higher-order corrections in $g$.

The IR asymptotic behaviour of the Green's functions
is determined by IR attractive fixed points of the corresponding RG equations. The coordinates of fixed points $g^*$, $u^*$ are found from the requirement that all the $\beta$ functions vanish simultaneously:
\begin{equation}
    \beta_g (g^*,u^*)=  \beta_u (g^*,u^*) = 0.
\end{equation}
The type of a fixed point is determined by the matrix of derivatives $\Omega_{ij} = \partial_i \beta_j (g^*)$ at the given point $g_i=\{g,u\}$: 
for an IR attractive point all the eigenvalues should have positive real parts.

Analysis of the expressions (\ref{Betau}) reveals four fixed points:

(i) Gaussian (free) fixed point:
\begin{equation}
    g^* = 0 \,, \quad u^* = 0;
\end{equation}

(ii) nontrivial fixed point:
\begin{equation}
    g^* = \frac{2(\varepsilon-\eta)^2}{\varepsilon-2\eta}\,, \quad 
    u^* = \frac{\varepsilon}{\varepsilon-2\eta}.
\end{equation}

The point (i) is IR attractive for $\varepsilon<0\,, \eta<0$, while
the point (ii) is IR attractive for $\varepsilon>0\,, \eta<\varepsilon/2$.

Two more points are found in the following way.
In order to explore the limiting case $u \to \infty$ with $g/u$ fixed, we have to pass to new variables: $g' \equiv g/u$ and $w \equiv 1/u$. For this case we obtain
\begin{eqnarray}
\beta_{g'} = g' \left[ \eta - \varepsilon + \frac{ g' }{w+1}\right]\,, \qquad
\beta_w = w \left[ \eta + g'\,\frac{w-1}{(w+1)^2}\right]\, .
\label{Betaw}    
\end{eqnarray}
Finding the zeros of the $\beta$ functions, we find two additional fixed points:

(iii) Gaussian (free) fixed point:
\begin{equation}
     g'^* = 0\,, \quad w^* = 0; 
\end{equation}

(iv) nontrivial fixed point:
\begin{equation}
     g'^* = \varepsilon - \eta\,, \quad w^* = 0.
\end{equation}

 The point (iii) is IR attractive if $\varepsilon>0\,, \varepsilon/2<\eta<\varepsilon$, and the point (iv) is IR attractive if $\varepsilon<0\,,\eta>0$ or $\varepsilon>0\,, \eta>\varepsilon$. 

\begin{figure}[h]
\centering
\includegraphics[width=0.6\linewidth]{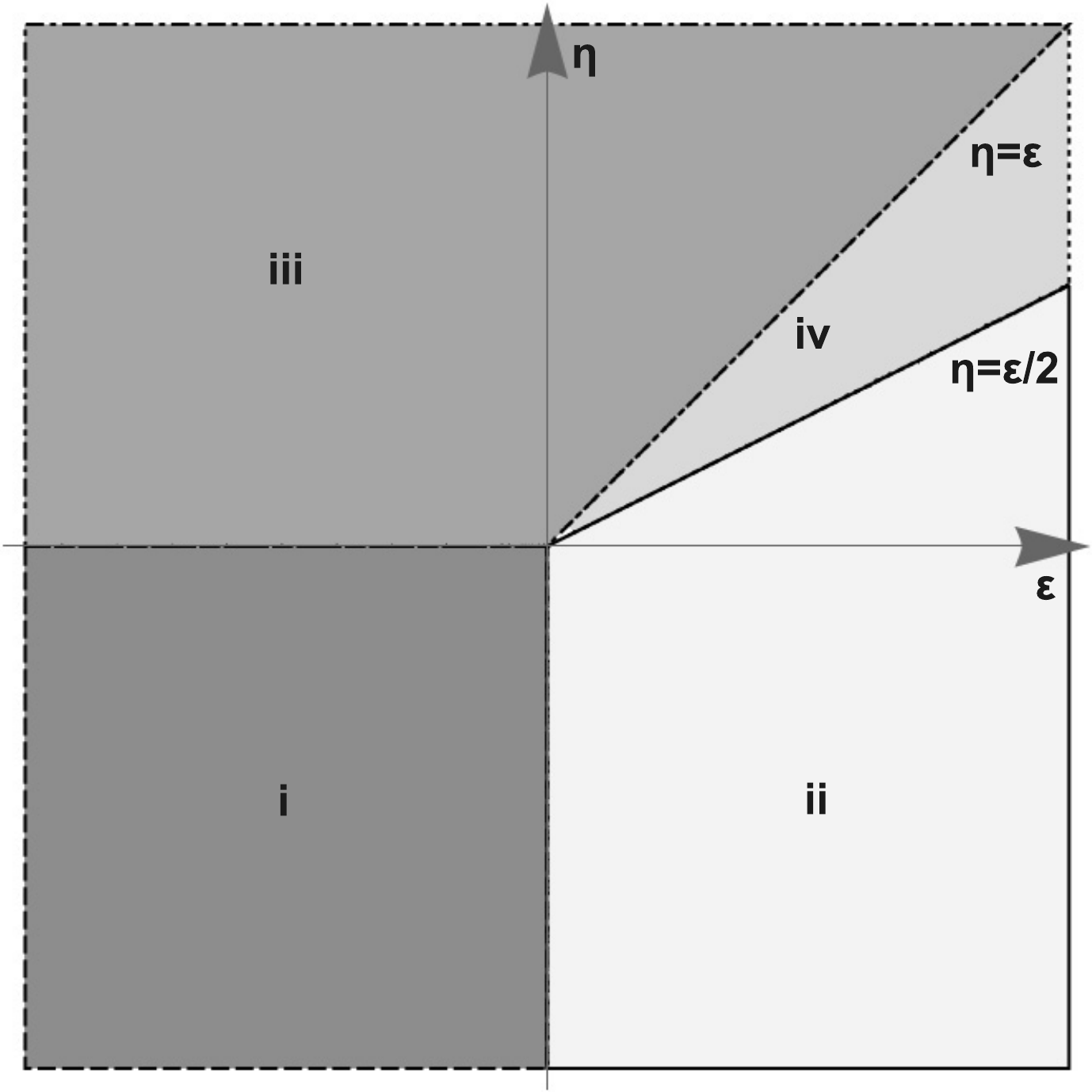}
\caption{Regions of stability of the fixed points (i)--(iv).}
\label{zone}
\end{figure}

 The general stability
 pattern of the fixed points in the $\varepsilon$--$\eta$ plane is shown in Fig.~\ref{zone}. 

 In the one-loop approximation, the regions of IR stability 
 for all the points are given by sectors that cover the full plane without gaps or overlaps between them.

Some remarks are in order. Clearly, the Gaussian points correspond to cases, in which the dynamics of the field $\theta$ is not affected by the
statistics of the height field $h$ (only in the leading order of the IR asymptotic behaviour!). In these cases, we deal with an ordinary random walk.

The point (iv) corresponds to the limiting case (\ref{Rapid}) when the field $h$, in comparison with $\theta$, behaves as if it was $\delta$-correlated in time.

However, we did not find a nontrivial point that would correspond to the frozen limit (\ref{Frozen}). This follows from the fact that the function
$\beta_g$ in (\ref{Betau}) becomes trivial for $u\to 0$: $\beta_g=-\varepsilon g$.
The similar triviality was observed earlier in models of diffusion in  time-independent potential vector fields where it was shown to be exact in all orders of perturbation theory \cite{Yudson,Juha}. Since those models have a close formal resemblance with the limit (\ref{Frozen}) of our model and its special case (\ref{EW}), (\ref{quench}), we believe that in the latter cases $\beta_g$ is also trivial exactly.

\section{Critical dimensions and scaling behaviour}
\label{Scaling}

Existence of IR attractive fixed points of the RG equations implies existence of the scaling behaviour of the correlation functions in the IR range.

In dynamical models, the critical dimension of any quantity $F$ (a field or a parameter) is given by the expression (see, e.g. Secs.~5.16 and 6.7 in \cite{Vasiliev} and Sec.~2.1 in \cite{RedBook}): 
\begin{equation}
\Delta_F = d^k_F + \Delta_{\omega} d^{\omega}_F + \gamma^*_F, \quad 
\Delta_{\omega} =2-\gamma_{\nu}^*
\label{DeltaF}
\end{equation}
(with the standard normalization convention that $\Delta_{\bf k}=-\Delta_{\bf x}=1$). Here and below $\gamma^*$ denotes the value of the anomalous dimension $\gamma$ at a fixed point.

For the Gaussian points (i) and (iii), one has 
\begin{eqnarray}
\Delta_{\theta'\theta}=d, \quad \Delta_{\omega}=2.
\label{RelationG}
\end{eqnarray}

For the fixed point (ii), one obtains the exact results from the relations (\ref{Gammas3}) and definitions (\ref{Gammas4}):
\begin{eqnarray}
\Delta_{\theta'\theta}=d, \quad \Delta_{\omega}=2-\eta.
\label{Relation2}
\end{eqnarray}

As already mentioned, the point (iv) corresponds to the 
limit (\ref{Rapid}), where the propagator $\langle hh \rangle_0$ becomes $\delta$-correlated in time.  As a result, closed circuits of retarded propagators $\langle \theta\theta' \rangle_0$ appear in almost all diagrams relevant for renormalization procedure and they therefore vanish. The only exception is the one-loop diagram contributing to $Z_1$.    
Thus, one has $Z_2=1$ identically, while $Z_1$ is given exactly by the one-loop expression, cf. the discussion of Kraichnan's rapid-change model of passive scalar advection \cite{AAV}. 
Then one readily derives exact expressions for the critical dimensions:  
\begin{eqnarray}
\Delta_{\theta'\theta}=d, \quad \Delta_{\omega}=2-\varepsilon+\eta.
\label{Relation4}
\end{eqnarray}


As an illustrative application, consider the mean-square distance  of a random walker on a rough surface. For such particle that started moving at $t=0$ from the origin ${\bf x}=0$, it is given by:
\begin{equation}
R^2(t) = \int\, d{\bf x}\, x^2 \langle \theta(t,{\bf x})
\theta'(0,{\bf 0}) \rangle,
\label{cloud}
\end{equation}
where $t>0$ is a later time and ${\bf x}$ is the corresponding current position.
Substituting the scaling representation for the linear response function
\begin{equation}
\langle \theta(t,{\bf x}) \theta'(0,{\bf 0})
\rangle \simeq r^{-\Delta_{\theta\theta'}}\,
F(tr^{-\Delta_{\omega}})
\label{response}
\end{equation}
gives:
 \begin{equation}
R^2(t) \propto t^{(d+2-\Delta_{\theta\theta'})/\Delta_{\omega}}.
\label{cloud9}
\end{equation}
Taking into account the exact relation $\Delta_{\theta'\theta}=d$, valid for all fixed points (i)-(iv), one arrives at the spreading law
\begin{equation}
R^2(t) \propto t^{2/\Delta_{\omega}},
\label{cloud10}
\end{equation}
with the exact expressions  $\Delta_{\omega}=2$ for the points (i), (iii), $\Delta_{\omega}=2-\eta$ for (ii) and $\Delta_{\omega}=2-\varepsilon+\eta$
for (iv).

\section{Conclusion}
\label{Conc}

We studied a model of a random walk of a particle on a rough fluctuating surface described by the Fokker--Planck equation for a particle in a constant gravitational field, while the surface was modelled by the (generalized) Edwards--Wilkinson model. The full stochastic problem (\ref{FPE}), (\ref{gravi}), (\ref{EWG}), (\ref{White2}) is mapped onto a multiplicatively renormalizable field theoretic model (\ref{Action}), (\ref{Actionh}).

The corresponding RG equations reveal two Gaussian (free) and two nontrivial fixed points, which means that the system exhibits various types of IR scaling behaviour (long times, large distances). Although the practical calculation is confined within the leading one-loop approximation, the main critical dimensions are found exactly.

As an illustrative example we considered the mean-square displacement of a walking particle (in other interpretation, the radius of particles' cloud). It shows that
the particle is not trapped in a finite area but travels all across the system with a spreading law similar to the ordinary random walk but, in general, with different exponents; see (\ref{cloud10}) and the text below.

As one can see, even a comparatively simple model demonstrates interesting types of IR behaviour. Thus, it is interesting to study more involved situations. There are several directions of possible generalization.

A linear stochastic equations like (\ref{EW}), (\ref{EWG})
(corresponding to Gaussian statistics for the height field) can be replaced by nonlinear models like the Kardar--Parisi--Zhang \cite{KPZ}
or Pavlik's \cite{Pavlik,AV} ones.

On some occasions, motion of a particle is not an ordinary random walk (\ref{RW})
but is described, e.g. by L\'{e}vy flights; see, e.g.~\cite{WalksX}. 
This possibility is supported by the ideas of self-organized criticality that the underlying surface evolves via avalanches~\cite{Bak3}, while the particle can ``glade'' upon the surface. If so, it is natural to replace the Laplace operator in the Fokker--Planck equation (\ref{FPE}) by a fractional derivative: $-\partial^2 \sim k^2 \to k^{2-\eta'}$ with a certain new exponent $\eta'$.

It is especially interesting to include anisotropy (as a consequence of an overall tilt of the surface). This can be done by describing the field $h$ by the
Pastor-Satorras--Rothman model for eroding landscape \cite{Pastor1,Pastor2}
or the Hwa--Kardar model of a running sandpile \cite{SOC1,SOC2}.

This work remains for the future and is partly in progress.

\section*{Acknowledgments}

The Authors are indebted to M.A.~Reiter for discussion.

The work of P.I.K. was supported by the Foundation for the Advancement of Theoretical Physics and Mathematics ``BASIS'', project 22-1-3-33-1 and by the Ministry of Science and Higher Education of the Russian Federation, agreement 075–15–2022–287.

\reftitle{References}

\end{document}